\documentclass[12pt]{article}%
\usepackage{amsmath}
\usepackage{amsfonts}
\usepackage{amssymb}
\usepackage{graphicx}%
\newtheorem{theorem}{Theorem}

\newenvironment{proof}[1][Proof]{\noindent\textbf{#1.} }{\ \rule{0.5em}{0.5em}}
\begin{document}

\title{On a general solution of the one-dimensional stationary Schr\"{o}dinger equation}
\author{Vladislav V. Kravchenko\\{\small Department of Mathematics, CINVESTAV del IPN, Unidad Quer\'{e}taro}\\{\small Libramiento Norponiente No. 2000, Fracc. Real de Juriquilla}\\{\small Queretaro, Qro. C.P. 76230 MEXICO}\\{\small e-mail: vkravchenko@qro.cinvestav.mx}}
\maketitle

\begin{abstract}
The general solution of the one-dimensional stationary Schr\"{o}dinger
equation in the form of a formal power series is considered. Its efficiency
for numerical analysis of initial value and boundary value problems is discussed.

\end{abstract}

Consider the equation%

\begin{equation}
(pu^{\prime})^{\prime}+qu=\omega^{2}u\label{onedimmaineq}%
\end{equation}
where we suppose that $p$, $q$ and $u$ are complex-valued functions of an
independent real variable $x\in\lbrack0,a]$, $\omega$ is an arbitrary complex
number. $p$ and $q$ are supposed to be such that there exists a solution
$g_{0}$ of the equation $(pg_{0}^{\prime})^{\prime}+qg_{0}=0$ on $(0,a)$ such
that $g_{0}\in C^{2}(0,a)$ together with $1/g_{0}$ are bounded on $[0,a]$ and
$p\in C^{1}(0,a)$ is a bounded nonvanishing function on $[0,a]$. Denote
$g=\sqrt{p}g_{0}$. In \cite{KrSolveSchr1} with the aid of pseudoanalytic
function theory \cite{Berskniga} the following result was obtained.

\begin{theorem}
\label{ThGenSolSturmLiouville}The general solution of (\ref{onedimmaineq}) has
the form
\begin{equation}
u=c_{1}u_{1}+c_{2}u_{2} \label{genmain}%
\end{equation}
where $c_{1}$ and $c_{2}$ are arbitrary complex constants, $u_{1}$ and $u_{2}$
are defined as follows
\begin{equation}
u_{1}=g_{0}%
{\displaystyle\sum\limits_{\text{even }n=0}^{\infty}}
\frac{\omega^{n}}{n!}\widetilde{X}^{(n)}\quad\text{and}\quad u_{2}=g_{0}%
{\displaystyle\sum\limits_{\text{odd }n=1}^{\infty}}
\frac{\omega^{n}}{n!}X^{(n)} \label{genmain1}%
\end{equation}
where $\widetilde{X}^{(n)}$ and $X^{(n)}$ are introduced by the following
recurrent equalities%
\begin{equation}
\widetilde{X}^{(0)}\equiv1,\quad X^{(0)}\equiv1, \label{Xgen1}%
\end{equation}
and for $n\in\mathbb{N}$,
\end{theorem}

\begin{equation}
\widetilde{X}^{(n)}(x)=\left\{
\begin{tabular}
[c]{ll}%
$n%
{\displaystyle\int\limits_{0}^{x}}
\widetilde{X}^{(n-1)}(\xi)g_{0}^{2}(\xi)d\xi$ & $\text{for an odd }n$\\
$n%
{\displaystyle\int\limits_{0}^{x}}
\widetilde{X}^{(n-1)}(\xi)g^{-2}(\xi)d\xi$ & $\text{for an even }n$%
\end{tabular}
\ \right.  \label{Xgen2}%
\end{equation}

\begin{equation}
X^{(n)}(x)=\left\{
\begin{tabular}
[c]{ll}%
$n%
{\displaystyle\int\limits_{0}^{x}}
X^{(n-1)}(\xi)g^{-2}(\xi)d\xi$ & $\text{for an odd }n$\\
$n%
{\displaystyle\int\limits_{0}^{x}}
X^{(n-1)}(\xi)g_{0}^{2}(\xi)d\xi$ & $\text{for an even }n$%
\end{tabular}
\ \ \ \ \right.  \label{Xgen3}%
\end{equation}
Another representation of the general solution of (\ref{onedimmaineq}) as a
formal power series has been known since quite long ago (see \cite[Theorem
1]{Trubowitz}) and used for studying qualitative properties of solutions. The
parameter $\omega$ participated in that representation in a very complicated
manner which made that form of a general solution too difficult to be used for
quantitative analysis of spectral and boundary value problems. To the
contrast, the solution (\ref{genmain}), (\ref{genmain1}) is a power series
with respect to $\omega$ which makes it really attractive for numerical
solution of spectral, initial value and boundary value problems.

The required for (\ref{genmain1}) particular solution $g_{0}$ can be
constructed in a similar way. By analogy with theorem
\ref{ThGenSolSturmLiouville} the following result can be obtained.

\begin{theorem}
\label{ThGenSolSchr}Suppose that $q\in C[0,a]$. The general solution of the
equation
\begin{equation}
-\frac{d^{2}u(x)}{dx^{2}}+q(x)u(x)=0 \label{main}%
\end{equation}
on $\left(  0,a\right)  $ has the form%
\begin{equation}
u=c_{1}u_{1}+c_{2}u_{2} \label{gensolSchr1}%
\end{equation}
where $c_{1}$ and $c_{2}$ are arbitrary constants; $u_{1}$, $u_{2}$ are
defined as follows%
\begin{equation}
u_{1}=%
{\displaystyle\sum\limits_{\text{even }n=0}^{\infty}}
\frac{\widetilde{X}^{(n)}}{n!}\qquad\text{and}\qquad u_{2}=%
{\displaystyle\sum\limits_{\text{odd }n=1}^{\infty}}
\frac{X^{(n)}}{n!}. \label{gensolSchr}%
\end{equation}
and $\widetilde{X}^{(n)}$, $X^{(n)}$ are introduced by the following recurrent
equalities
\begin{equation}
\widetilde{X}^{(0)}\equiv1,\quad X^{(0)}\equiv1, \label{X1}%
\end{equation}
and for $n\in\mathbb{N}$,
\end{theorem}

\begin{equation}
\widetilde{X}^{(n)}(x)=\left\{
\begin{tabular}
[c]{ll}%
$n%
{\displaystyle\int\limits_{0}^{x}}
\widetilde{X}^{(n-1)}(\xi)d\xi$ & $\text{for an even }n$\\
$n%
{\displaystyle\int\limits_{0}^{x}}
\widetilde{X}^{(n-1)}(\xi)q(\xi)d\xi$ & $\text{for an odd }n$%
\end{tabular}
\ \ \right.  \label{X2}%
\end{equation}

\begin{equation}
X^{(n)}(x)=\left\{
\begin{tabular}
[c]{ll}%
$n%
{\displaystyle\int\limits_{0}^{x}}
X^{(n-1)}(\xi)q(\xi)d\xi$ & $\text{for an even }n$\\
$n%
{\displaystyle\int\limits_{0}^{x}}
X^{(n-1)}(\xi)d\xi$ & $\text{for an odd }n$%
\end{tabular}
\ \ \ \right.  \label{X3}%
\end{equation}

\begin{proof}
First of all let us verify that both series are uniformly convergent on the
considered interval. For this purpose we notice that for an even $n$,
\[
\left\vert \widetilde{X}^{(n)}(x)\right\vert \leq\left(  \max_{x\in
\lbrack0,x]}\left\vert q(x)\right\vert \right)  ^{n/2}x^{n}\leq\left(
\max_{x\in\lbrack0,a]}\left\vert q(x)\right\vert \right)  ^{n/2}a^{n}.
\]
Thus, the members of the series in $u_{1}$ can be estimated by constants:
$\frac{\left\vert \widetilde{X}^{(n)}(x)\right\vert }{n!}\leq\frac{\left(
\max_{x\in\lbrack0,a]}\left\vert q(x)\right\vert \right)  ^{n/2}a^{n}}{n!}$
for any $x\in\lbrack0,a]$ and the series $\frac{c^{n}}{n!}$ converges where
$c=\left(  \max_{x\in\lbrack0,a]}\left\vert q(x)\right\vert \right)  ^{1/2}a.$
Then by the Weierstrass theorem the series in $u_{1}$ is uniformly convergent.
The uniform convergence of the series in $u_{2}$ (as well as of the series of
derivatives) can be shown by analogy.

Now we can apply the operator $\frac{d^{2}}{dx^{2}}$ to each of the series.
Note that for an even $n>0$,
\[
\frac{d^{2}}{dx^{2}}\widetilde{X}^{(n)}=n\frac{d}{dx}\widetilde{X}%
^{(n-1)}=(n-1)nq\widetilde{X}^{(n-2)}.
\]
Thus,
\[
\frac{d^{2}}{dx^{2}}u_{1}=q%
{\displaystyle\sum\limits_{\text{even }n=2}^{\infty}}
\frac{\widetilde{X}^{(n-2)}}{\left(  n-2\right)  !}=q%
{\displaystyle\sum\limits_{\text{even }n=0}^{\infty}}
\frac{\widetilde{X}^{(n)}}{n!}=qu_{1}.
\]
For an odd $n>1$ (obviously, $\frac{d^{2}X^{(1)}}{dx^{2}}=0$) we have
\[
\frac{d^{2}}{dx^{2}}X^{(n)}=n\frac{d}{dx}X^{(n-1)}=(n-1)nqX^{(n-2)}.
\]
And consequently%
\[
\frac{d^{2}}{dx^{2}}u_{2}=q%
{\displaystyle\sum\limits_{\text{odd }n=3}^{\infty}}
\frac{X^{(n-2)}}{\left(  n-2\right)  !}=q%
{\displaystyle\sum\limits_{\text{odd }n=1}^{\infty}}
\frac{X^{(n)}}{n!}=qu_{2}.
\]

Thus, $u_{1}$ and $u_{2}$ are solutions of (\ref{main}). Last step is to
verify that their Wronskian is different from zero at least at one point. It
is easy to see that the Wronskian has the form%
\[
\left(
{\displaystyle\sum\limits_{\text{even }n=0}^{\infty}}
\frac{\widetilde{X}^{(n)}}{n!}\right)  \left(
{\displaystyle\sum\limits_{\text{even }m=0}^{\infty}}
\frac{X^{(m)}}{m!}\right)  -\left(
{\displaystyle\sum\limits_{\text{odd }n=1}^{\infty}}
\frac{\widetilde{X}^{(n)}}{n!}\right)  \left(
{\displaystyle\sum\limits_{\text{odd }m=1}^{\infty}}
\frac{X^{(m)}}{m!}\right)  .
\]
At the point zero all $X^{(m)}$ and $\widetilde{X}^{(n)}$ vanish except for
$X^{(0)}$ and $\widetilde{X}^{(0)}$. Thus the Wronskian is equal to $1$ at
$x=0$ and hence the functions $u_{1}$, $u_{2}$ are linearly independent that
finishes the proof.
\end{proof}

This theorem in different, quite more difficult notations was known already at
the beginning of the last century (see \cite{Weyl}) and was also used for
qualitative analysis of solutions of (\ref{main}). Nevertheless here we want
to emphasize the extreme usefulness of theorem \ref{ThGenSolSchr} as well as
of theorem \ref{ThGenSolSturmLiouville} for numerical solution of initial
value and boundary value problems for second-order linear differential
equations, which escaped attention of researchers working in numerical
solution of ordinary differential equations. First of all, let us notice that
in (\ref{gensolSchr1}) $c_{1}=u(0)$ and $c_{2}=u^{\prime}(0)$. That is the
representation of a general solution (\ref{gensolSchr1}), (\ref{gensolSchr})
is very convenient for solving initial value problems. We remind that solution
of boundary value problems for (\ref{main}) reduces to solving a couple of
initial value problems (see, e.g., \cite{Stan}), so this property of the
representation (\ref{gensolSchr1}), (\ref{gensolSchr}) is well suited for
solving boundary value problems as well.

Moreover, note that very often, e.g., in electromagnetic theory (see,
\cite{Wait}) it is necessary to solve the equation
\begin{equation}
-\frac{d^{2}u(x)}{dx^{2}}+\omega^{2}q(x)u(x)=0 \label{emSchr}%
\end{equation}
for different values of the complex constant $\omega^{2}$. According to
theorem \ref{ThGenSolSchr} its general solution can be represented as follows%
\[
u=c_{1}%
{\displaystyle\sum\limits_{\text{even }n=0}^{\infty}}
\frac{\omega^{n}\widetilde{X}^{(n)}}{n!}+c_{2}%
{\displaystyle\sum\limits_{\text{odd }n=1}^{\infty}}
\frac{\omega^{n-1}X^{(n)}}{n!}%
\]
with $X^{(n)}$ and $\widetilde{X}^{(n)}$ defined by (\ref{X1})-(\ref{X3}).
Thus, once $X^{(n)}$ and $\widetilde{X}^{(n)}$ up to a certain order $N$ are
calculated, an approximate solution of (\ref{emSchr}) is just a polynomial in
$\omega$ with calculated coefficients $X^{(n)}$ and $\widetilde{X}^{(n)}$.
This observation is also valid in the case of the solution (\ref{genmain}),
(\ref{genmain1}) of equation (\ref{onedimmaineq}). This property is very
useful for numerical solution of corresponding spectral problems which then
reduces to finding zeros of polynomials with respect to $\omega$. In the
present work we are more interested in studying the convergence and accuracy
of the numerical method based on representations of the form
(\ref{gensolSchr1}), (\ref{gensolSchr}) in comparison with known standard algorithms.

An important feature of the representation (\ref{gensolSchr1}),
(\ref{gensolSchr}) is that it is well suited for symbolic calculations in
principle in a general case. The coefficient $q$ can be interpolated
arbitrarily accurately by means of a polynomial or splines and then all
integrations in (\ref{X2}) and (\ref{X3}) can be done symbolically in a
package like Mathematica or Maple. In the present work we made use of Matlab 7
and compared our results with standard Matlab ODE solvers \cite{Ashino},
\cite{Shampine}, especially with ode45 which in the considered examples gave
always better results than other similar programs.

Consider the following initial value problem for (\ref{main}): $q\equiv-c^{2}%
$, $u(0)=1$, $u^{\prime}(0)=-1$ on the interval $(0,1)$. For $c=1$ the
absolute error of the result calculated by ode45 (with an optimal tolerance
chosen) was of order $10^{-9}$ and the relative error was of order $10^{-6}$
meanwhile the absolute error of the result calculated with the aid of theorem
\ref{ThGenSolSchr} with $N$ (the number of formal powers in the truncated
series (\ref{gensolSchr})) from $55$ to $58$ was of order $10^{-16}$ and the
relative error was of order $10^{-14}$. Taking $c=10$ under the same
conditions the absolute and the relative errors of ode45 were of order
$10^{-6}$ and $10^{-5}$ respectively meanwhile our algorithm based on theorem
\ref{ThGenSolSchr} gave values of order $10^{-12}$ in both cases.

For the initial value problem: $q\equiv c^{2}$, $u(0)=1$, $u^{\prime}(0)=-1$
on the interval $(0,1)$ in the case $c=1$ the absolute and the relative errors
of ode45 were of order $10^{-8}$ meanwhile in our method this value was of
order $10^{-15}$ already for $N=50$. For $c=10$ the absolute and the relative
errors of ode45 were of order $10^{-3}$ and $10^{-7}$ respectively and in the
case of our method these values were of order $10^{-11}$ and $10^{-14}$ for
$N=50$.

Consider another example. Let $q(x)=c^{2}x^{2}+c$. In this case the general
solution of (\ref{main}) has the form
\[
u(x)=e^{cx^{2}/2}\left(  c_{1}+c_{2}\int_{0}^{x}e^{-ct^{2}}dt\right)  .
\]
Take the same initial conditions as before, $u(0)=1$, $u^{\prime}(0)=-1$. Then
meanwhile for $c=1$ the absolute and the relative error of ode45 was of order
$10^{-8}$ and for $c=30$ the absolute error was $0.28$ and the relative error
was of order $10^{-6}$, our algorithm ($N=58$) gave the absolute and relative
errors of order $10^{-15}$ for $c=1$ and the absolute and relative errors of
order $10^{-9}$ and $10^{-15}$ respectively for $c=30.$ All calculations were
performed on a usual PC with the aid of Matlab 7.

The results of our numerical experiments show that in fact theorem
\ref{ThGenSolSturmLiouville} and theorem \ref{ThGenSolSchr} offer a new
powerful method for numerical solution of initial value and boundary value
problems for linear ordinary differential second-order equations. Numerical
calculation of integrals involved in (\ref{Xgen2}), (\ref{Xgen3}) and in
(\ref{X2}), (\ref{X3}) does not represent any considerable difficulty and can
be done with a remarkable accuracy.

\end{document}